\documentclass[twocolumn,showpacs,preprintnumbers,amsmath,amssymb]{revtex4}
\usepackage[dvips]{graphicx}%Include figure files
\usepackage{dcolumn}% Align table columns on decimal point
\usepackage{bm}% bold math
\usepackage{hyperref}
\hypersetup{colorlinks=blue}

\begin{document}

\preprint{APS/123-QED}

\title{Unusual Quasiparticle Tunneling in High-$T_c$ Cuprate Superconductors: Evidence for the BCS and Polaronic Multi-Gap Effects on Tunneling Spectra}

\author{S. Dzhumanov}
\email{dzhumanov@rambler.ru} \affiliation{%
Institute of Nuclear Physics, Uzbek Academy of Sciences, 100214,
Ulughbek, Tashkent, Uzbekistan}
\author{O.K. Ganiev}
\affiliation{%
Institute of Nuclear Physics, Uzbek Academy of Sciences, 100214,
Ulughbek, Tashkent, Uzbekistan}
\author{Sh.S. Djumanov}
\affiliation{%
Institute of Nuclear Physics, Uzbek Academy of Sciences, 100214,
Ulughbek, Tashkent, Uzbekistan}
%%\date{\today}% It is always \today, today,

\begin{abstract}
We propose a model of quasiparticle tunneling across the high-$T_c$
superconductor-insulator-normal metal junction based on the
different mechanisms for tunneling of electrons at positive bias and
dissociating polaronic Cooper pairs and large polarons at negative
bias, and the gap inhomogeneity (i.e., multi-gap) picture. We show
that the main features of the tunneling spectra such as low-bias U-
and V-shaped features, asymmetry and high-bias dip-hump features,
their temperature and doping dependences, and shoulders inside  the
conductance peaks observed in high-$T_c$ cuprates arise naturally
from the model. The experimental tunneling spectra of
$\rm{Ba_2Sr_2CaCu_2O_{8+\delta}}$ are fitted quite well by taking
into account the distribution of BCS and polaronic gap values.
\end{abstract}

%\keywords{}%Use showkeys class option if keyword
%                              %display desired
\maketitle

In conventional superconductors, superconductivity arises from the
binding of electrons into Cooper pairs and the BCS gap opening in
the electronic excitation spectrum at the superconducting (SC)
transition temperature $T_c$ serves as the SC order parameter. In
contrast, in high-$T_c$ cuprate superconductors the precursor Cooper
pairing of carriers occurs in the normal state and the BCS-like gap
is manifested as a pseudogap (PG) opening at a temperature $T^*$
higher than $T_c$ at which the Cooper pairs condense into a
superfluid (SF) Bose-liquid state \cite{1,2}. As argued in Refs.
\cite{1,2,3}, in these materials the Cooper pairing is only a
necessary but not a sufficient condition for the occurrence of
superconductivity and the BCS-like gap might be different from the
SC order parameter appearing below $T_c$. Other alternative PG
scenarios were also proposed (for a review, see, e.g., Refs.
\cite{4,5,6}), among which the SC fluctuation scenario suggests that
the PG appearing below $T^*$ is related to superconductivity as a
form of precursor pairing and evolves smoothly into the SC gap at
$T_c$ \cite{7}. It is still highly debated whether the PG observed
by various experimental techniques originates from SC fluctuations
above $T_c$ \cite{7,8,9} or from the precursor non-SC Cooper pairing
below $T^*$ \cite{1,2,3} or from some other effects (see Refs.
\cite{10,11,12}).

Scanning tunneling microscopy and spectroscopy (STM and STS)
\cite{13,14,15,16,17} and angle-resolved photoemission spectroscopy
(ARPES) \cite{18,19,20,21} have made significant progress in the
studies of PG phenomena in high-$T_c$ cuprates and other materials.
Progressive investigations of tunneling and ARPES spectra have
provided important information on single-particle excitation gaps in
the cuprates. The STM and STS techniques are very sensitive to the
quasiparticle density of states (DOS), with the unique capability to
measure any excitation gaps at the Fermi energy, and to the
electronic (or gap) inhomogeneities that are intrinsic to the
cuprates \cite{15,16,17,22,23,24}. These tunneling studies of the
cuprate superconductors have revealed a rich variety of tunneling
spectra (differential conductance-voltage ($dI/dV-V$)
characteristics) of high-$T_c$ cuprate superconductor
(HTSC)-insulator (I)-normal metal (N) (called SIN) junctions. Two
distinctive features of the $dI/dV$ spectra are nearly U- and
V-shaped characteristics observed in tunneling experiments on
high-$T_c$ cuprates \cite{15,24,25,26,27,28,29}. Other distinctive
tunneling features systematically observed in the $dI/dV-V$
characteristics of SIN junctions are:(i) asymmetric conductance
peaks \cite{26,27,28,29,30}, (ii) dip-hump structures appearing
outside the conductance peak on the negative bias side
\cite{13,27,28,30}, (iii) suppression of the peak on the negative
bias side with increasing temperature and its vanishing somewhat
below $T_c$ or near $T_c$, leaving the hump feature (i.e., linearly
increasing conductance) and the second conductance peak (on the
positive bias side) \cite{15,27,28}, and (iv) shoulders inside the
conductance peaks \cite{22,23,24}. Similar peak-dip-hump feature and
its persistence above $T_c$ were also observed in ARPES spectra
\cite{18,31}.

The unusual features of the tunneling spectra of high-$T_c$ cuprates
are neither expected within the simple s-wave BCS model nor within
the BCS models based on the d-wave gap symmetry. For example, the
tunneling spectra showing more flatter, BCS-like feature is
difficult to reproduce in the d-wave model. While the other type of
tunneling spectra showing more V-shaped feature might be expected
either within the d-wave gap model (which fails to reproduce
quantitatively the conductance peak height and shape
\cite{13,15,24}) or within the s-wave multi-gap model proposed in
the present work. The origins of the peak-dip-hump feature and the
asymmetry of the conductance peaks have also been the subject of
controversy. These features of the tunneling spectra have been
attributed either to extrinsic (band-structure) effects (i.e., the
van Hove singularity and bilayer splitting) \cite{32,33} or to
intrinsic effects such as particle-hole asymmetry \cite{34,35} and
strong coupling effects (which originate from the coupling to a
phonon mode \cite{36} or to a collective electronic mode \cite{34}).
Although several theoretical models were successful in reproducing
some tunneling spectra with the peak-dip-hump features and
asymmetric or nearly symmetric peaks observed in cuprates, the
well-established experimental tunneling spectra with different
peak-dip-hump features \cite{13,15} (e.g., high-bias conductances,
which are nearly flat, linearly increasing at negative bias or
decreasing at positive bias, temperature- and doping-dependent
peaks, dip-hump features and asymmetry of the conductance peaks)
characteristic of high-$T_c$ cuprates were not explained yet. The
quantitative fits of the experimental spectra of
$\rm{Ba_2Sr_2CaCu_2O_{8+\delta}}$ (Bi-2212) to a BCS d-wave gap and
a T-independent gap functions \cite{15} have been unsuccessful in
reproducing the asymmetry (which is opposite to that of the observed
tunneling spectra) and the temperature-dependent conductance curves.
For underdoped Bi-2212, the asymmetry of the conductance peaks and
its doping dependence found in Ref. \cite{30} are also inconsistent
(i.e., in contrast) with other experimental data \cite{15,27,28}.
Further, STM and STS studies have shown that the electronic system
of Bi-2212 is inhomogeneous and the gap distribution (i.e.,
inhomogeneity) has a strong effect on the tunneling spectra
\cite{22,23,24,37}. It is necessary for any theoretical model to
explain not only the asymmetry and peak-dip-hump features in
tunneling conductance but also their evolution with temperature and
doping, a flat (i.e., U- shaped) and more V- shaped subgap
conductances, the gap-inhomogeneity-induced shoulders inside the
conductance peaks that are observed recently by STM in tunneling
spectra of Bi-2212 \cite{23,24}.

In this paper, we propose a simple and quite effective model of
quasiparticle tunneling based on the different mechanisms for
tunneling of charge carriers across the SIN junction at negative and
positive biases and the multi-gap (i.e., gap inhomogeneity) picture.
The proposed model reproduces the well-established experimental
tunneling spectra of high-$T_c$ cuprates and their nearly U and
V-shaped features, asymmetry, peak-dip-hump structure and
shoulder-like features inside the conductance peaks. We focus on the
Bi-2212 system (which has been well studied experimentally) and show
that the main experimental features of the tunneling spectra and
their temperature and doping dependences can be well reproduced by
using a BCS DOS at positive bias voltages ($V>0$) and the combined
BCS DOS and quasi-free-state DOS (originating from the dissociation
of large polarons) at negative bias voltages ($V<0$), and taking
into account the distribution of BCS and polaronic gap values. There
is now ample reason to believe that the electron-phonon interaction
in cuprates is strong enough and the relevant charge carriers in
these systems are large polarons \cite{3,38,39}. Therefore, the
precursor Cooper pairing of large polarons may occur above $T_c$
with opening the energy gap $\Delta$ in their excitation spectrum
\cite{1,2,3}. As argued in Ref. \cite{40}, the binding energies of
large polarons $\Delta_p$ and Cooper-like large polaron pairs are
manifested as the two distinct non-SC gaps in high-$T_c$ cuprates,
one a temperature independent PG and the other a BCS-like gap. Two
gap-like features observed in ARPES and tunneling experiments are
often misinterpreted as the coexisting PG and SC gap.

{\it{Unconventional SIN tunneling.}} -- We consider the model which
describes two specific mechanisms for quasiparticle tunneling across
the SIN junction at $V<0$ and $V>0$, and explains the asymmetry of
the tunneling current taking into account the different tunneling
DOS existing in these cases. The first mechanism describes the
$S\rightarrow N$ tunneling processes associated with the
dissociation of Cooper-like polaron pairs and large polarons at
$V<0$. In this case the Cooper pair dissociates into an electron in
a normal metal and a polaron in a polaron band of the HTSC. This
$S\rightarrow N$ tunneling is allowed only at $|eV|>\Delta$. The
dissociation of large polaron occurs at $|eV|>\Delta_p$ and the
carrier released from the polaron potential well can tunnel from the
quasi-free state into the free states of the normal metal. Such a
$S\rightarrow N$ transition gives an additional contribution to the
tunneling current. The other mechanism describes the electron
tunneling from the normal metal to the BCS-like quasiparticle states
in HTSC at $V>0$, while the quasi-free states appearing only at the
polaron dissociation are absent. Therefore, at $V>0$ the tunneling
current across SIN junction is proportional to the BCS DOS of HTSC
given by
\begin{eqnarray}\label{Eq.1}
D_{BCS}(E,\Delta)= \left\{ \begin{array}{ll}
D(\varepsilon_F)\frac{|E|}{\sqrt{E^2-\Delta^2}}& \textrm{for}\:
|E|>\Delta,\\
0 & \textrm{for}\:|E|<\Delta,
\end{array} \right.
\end{eqnarray}
where
$D(\varepsilon_F)=m_p^{3/2}\sqrt{\varepsilon_F}/\sqrt{2}\pi^2\hbar^3$
is the normal state DOS, $m_p$ and $\varepsilon_F$ are the mass and
Fermi energy of large polarons.

In the case $V<0$, the total current is the sum of two tunneling
currents and proportional to the square of the tunneling matrix
element, $|M|^2$ \cite{41}, the $D_{BCS}(E,\Delta)$ and the
quasi-free state DOS. This current flows from HTSC to normal metal
at the dissociation of Cooper pairs and large polarons. In HTSC, the
quasi-free carriers appearing at the dissociation of large polarons
have the effective mass $m^*$ and energy
$E=\Delta_p+\hbar^2k^2/2m^*$. Then the quasi-free state DOS is
defined as
\begin{eqnarray}\label{Eq.2}
D_f(E,\Delta_p)= \left\{ \begin{array}{ll}
D(\varepsilon^f_F)\sqrt{(|E|-\Delta_p)/\varepsilon_F^f}&
\textrm{for}\:
|E|>\Delta_p,\\
0 & \textrm{for}\:|E|<\Delta_p,
\end{array} \right.
\end{eqnarray}
where $\varepsilon^f_F$ is the Fermi energy of quasi-free carriers,
$D(\varepsilon^f_F)=(m^*)^{3/2}\sqrt{\varepsilon^f_F}/\sqrt{2}\pi^2\hbar^3$.
For the normal metal, the DOS at the Fermi energy $E_F$ is
independent of energy $E$, i.e., $D(E)\simeq D(E_F)$. Thus, at $V>0$
the tunneling current from the normal metal to HTSC is
\begin{eqnarray}\label{Eq.3}
&&I_{N\rightarrow
S}(V)=C|M|^2D(E_F)D(\varepsilon_F)\nonumber\\
&&\times\int\limits_{-\infty}^{+\infty}\frac{|E+eV|}{\sqrt{(E+eV)^2-\Delta^2}}\left[f(E)-f(E+eV)\right]dE\nonumber\\
&&=\frac{G}{e}\int\limits_{-\infty}^{+\infty}\frac{|\varepsilon|}{\sqrt{\varepsilon^2-\Delta^2}}\left[f(\varepsilon
-eV)-f(\varepsilon)\right]d\varepsilon,
\end{eqnarray}
where $G=e C|M|^2D(E_F)D(\varepsilon_F)$, $C$ is a constant,
$f(\varepsilon)$ is the Fermi function, $\varepsilon=E+eV$. The
differential conductance, $dI_{N\rightarrow S}/dV$ is then given by
\begin{eqnarray}\label{Eq.4}
dI_{N\rightarrow S}/dV=G(A_1(\Delta_T,a_V)+A_2(\Delta_T,a_V)),
\end{eqnarray}
where \vspace{-0.3cm}
\begin{eqnarray*}
A_1(\Delta_T,a_V)=\int\limits_{\Delta_T}^{+\infty}\frac{x
\exp[-x-a_V]dx}{\sqrt{x^2-\Delta_T^2}(\exp[-x-a_V]+1)^2},
\end{eqnarray*}
\vspace{-0.6cm}
\begin{eqnarray*}
A_2(\Delta_T,a_V)=\int\limits_{\Delta_T}^{+\infty}\frac{x
\exp[x-a_V]dx}{\sqrt{x^2-\Delta_T^2}(\exp[x-a_V]+1)^2},
\end{eqnarray*}
$ x=\varepsilon/k_B T,$ $a_V=eV/k_B T.$
\begin{figure}
\includegraphics[width=0.4\textwidth]{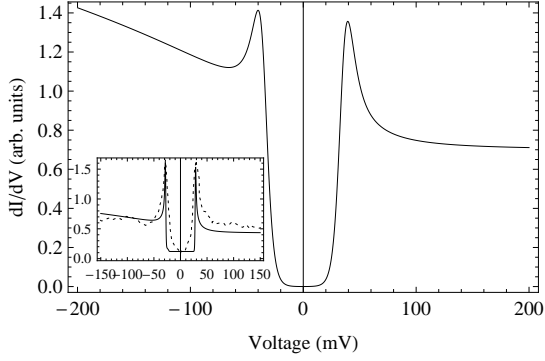}
\caption{\label{fig1:epsart}\footnotesize Main panel: SIN tunneling
conductance for T=40 K calculated using the one-gap model with
single $s$-wave BCS gap $\Delta$=35 meV and single polaronic gap
$\Delta_p$=40 meV, exhibiting U-shaped feature at low-bias. Inset:
comparison of the model ($\Delta$=28 meV and $\Delta_p$=22 meV,
solid line) with optimally doped Bi-2212 ($T_c$=92 K) tunneling data
at 4.8 K (dashed line) \cite{13}.}
\end{figure}
At negative bias voltages $V<0$, the tunneling current and
differential conductance are given by
\begin{eqnarray}\label{Eq.5}
I_{S\rightarrow
N}=\frac{G}{e}\left\{\int\limits_{-\infty}^{+\infty}\frac{|\varepsilon|d\varepsilon}{\sqrt{\varepsilon^2-\Delta^2}}[f(\varepsilon)-f(\varepsilon+eV)]\right.\nonumber\\
\left.+\frac{D(\varepsilon_F^f)}{D(\varepsilon_F)\sqrt{\varepsilon_F}}\int\limits_{-\infty}^{+\infty}\sqrt{|\varepsilon|-\Delta_p}[f(\varepsilon)-f(\varepsilon+eV)]d\varepsilon
\right\},
\end{eqnarray}
and
\begin{eqnarray}\label{Eq.4}
\frac{dI_{S\rightarrow
N}}{dV}=G\left\{A_1(\Delta_T,-a_V)+A_2(\Delta_T,-a_V)\right.\nonumber\\
\left.+a_F(T)[B_1(\Delta_p^*,a_V)+B_2(\Delta_p^*,a_V)]\right\},
\end{eqnarray}
\begin{figure}
\includegraphics[width=0.34\textwidth]{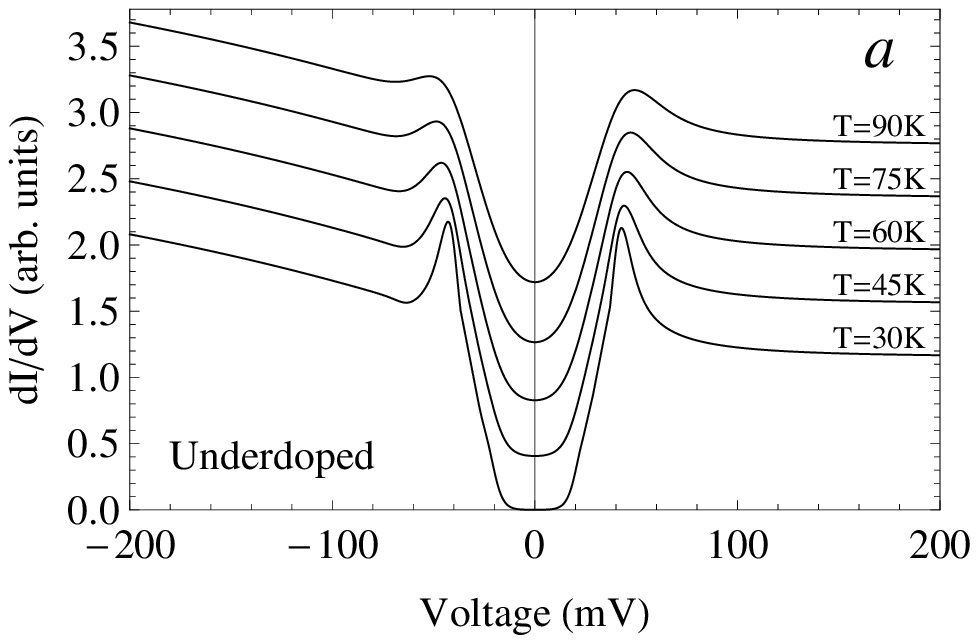}
\includegraphics[width=0.34\textwidth]{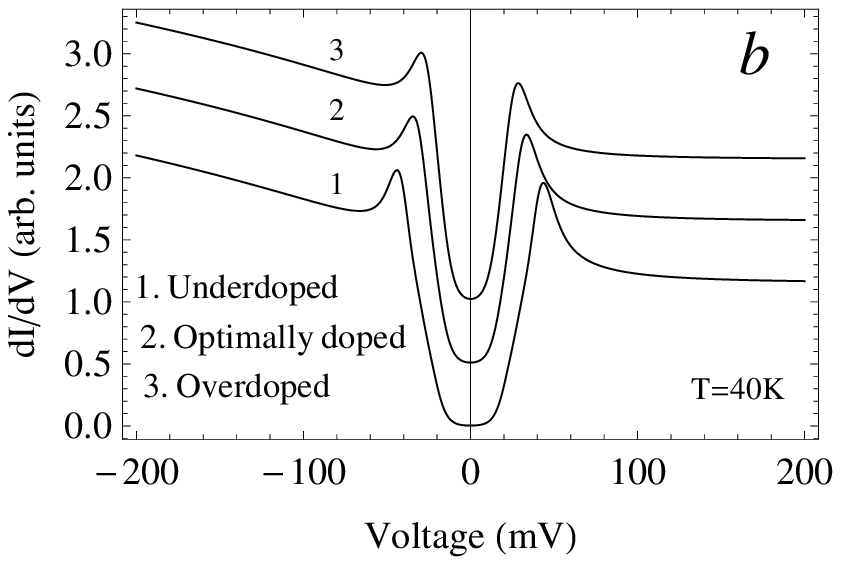}
\caption{\label{fig2:epsart} \footnotesize Tunneling conductance as
a function of temperature (a) and doping (b), calculated using the
three-gap model. In (a) the gap values are: $\Delta$=40, 30 and 22
meV, $\Delta_{p}$=65, 47 and 31 meV. In (b) $T$=40 K and set of gap
values: $\Delta$=40, 30 and 24 meV, $\Delta_{p}$=52, 38 and 28 meV
for curve 1; $\Delta$=30, 24 and 18 meV, $\Delta_{p}$=33, 28 and 22
meV for curve 2; and $\Delta$=25, 20 and 16 meV, $\Delta_{p}$=21, 17
and 14 meV for curve 3.}
\end{figure}
where $\varepsilon=E-eV$, \vspace{-0.3cm}
\begin{eqnarray*}
B_1(\Delta_p^*,a_V)=\int\limits_{\Delta_p^*}^{\infty}\sqrt{|x|-\Delta_p^*}\frac{\exp[x+a_V]dx}{(\exp[x+a_V]+1)^2},
\end{eqnarray*}
\vspace{-0.6cm}
\begin{eqnarray*}
B_2(\Delta_p^*,a_V)=\int\limits_{\Delta_p^*}^{\infty}\sqrt{|x|-\Delta_p^*}\frac{\exp[-x+a_V]dx}{(\exp[-x+a_V]+1)^2},
\end{eqnarray*}
\vspace{-0.6cm}
\begin{eqnarray*}
a_F(T)=[D(\varepsilon_F^f)/D(\varepsilon_F)]\sqrt{k_BT/\varepsilon_F},\quad
\Delta_p^*=\Delta_p/k_BT.
\end{eqnarray*}
The SIN tunneling conductance curve calculated at $T$=30 K for the
single-gap case (concerning both the polaronic gap $\Delta_p$ and
the $s$-wave BCS gap $\Delta$) is shown in Fig.1(main panel). In
this simple model, the absence of gap distribution would lead to the
U-shaped spectral behavior at low bias and such a more flatter
subgap conductance would be expected for homogeneous high-$T_c$
cuprates. As can be seen in Fig.1, there are dip-hump feature and
asymmetric peaks, with the higher peak in the negative bias voltage.
The model is compared with one of the best tunneling spectra
measured at $T\simeq4.8 K$ in Bi-2212 \cite{13}, as shown in the
inset of Fig.1.

{\it Multi-gap model.} -- One can expect that the electronic
inhomogeneity in HTSC may produce regions with a distribution of gap
amplitudes $(\Delta(i)$ and $\Delta_p(i))$ and variation in the
local DOS. Recent STM and STS experiments on Bi-2212 and other
high-$T_c$ systems indicate that the gap inhomogeneities commonly
exist in these materials regardless of doping level
\cite{22,23,24,37}. Therefore, in order to reproduce the main
features of the tunneling spectra of high-$T_c$ cuprates, we have to
consider the multi-gap case and the multi-channel tunneling
processes, which contribute to the total tunneling current. At
positive bias voltages $V>0$, the tunneling of electrons from the
normal metal into many regions of HTSC with different BCS DOS takes
place and the resulting conductance is
\begin{eqnarray}\label{Eq.7}
\frac{dI_{N\rightarrow S}}{dV}=\sum_{i}
G_i[A_{1i}(\Delta_T(i),a_V)+A_{2i}(\Delta_T(i),a_V)].
\end{eqnarray}
In the case $V<0$, the total current is the sum of tunneling
currents from many areas of HTSC with different local DOS
($D_{BCS}(E,\Delta(i))$ and $D_f(E,\Delta_p(i))$) to the normal
metal. Then the resulting conductance is
\begin{eqnarray}\label{Eq.8}
\frac{dI_{S\rightarrow
N}}{dV}=\sum_{i}G_i\{A_{1i}(\Delta_T(i),-a_V)+A_{2i}(\Delta_T(i),-a_V)\nonumber\\
+a_{Fi}(T)[B_{1i}(\Delta_p^*(i),a_V)+B_{2i}(\Delta_p^*(i),a_V)]\}.
\end{eqnarray}

In such a multi-gap model, the tunneling spectra exhibit a more
V-shaped behavior at low bias, the peak-dip-hump feature at negative
bias and the asymmetry of the conductance peaks. With increasing
temperature, the dip and peak on the negative bias side gradually
disappear (see Fig.2{\it a}), leaving the hump feature and the
second conductance peak (on the positive bias side), as observed in
tunneling experiments \cite{15}. Figure 2{\it b} shows that the
conductance peaks become more asymmetric with increasing doping, as
seen in experiments \cite{27,28}.

{\it Comparison with the experiment.} --The parameters entering into
Eqs.(\ref{Eq.7}) and (\ref{Eq.8}) can be varied to fit experimental
data. The comparison of the theoretical results with the different
experimental data on underdoped, slightly underdoped and overdoped
Bi-2212 is presented in Fig.3.
\begin{figure}
\includegraphics[width=0.46\textwidth]{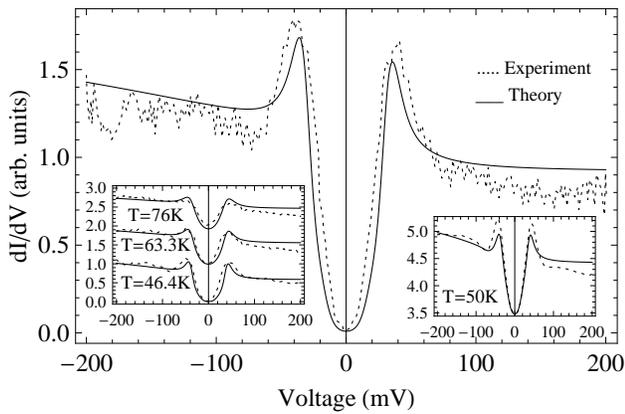}
\caption{\label{fig3:epsart}\footnotesize Main panel: SIN tunneling
spectrum measured on overdoped Bi-2212 at 43.1 K \cite{27} fitted by
using two-gap model, with $\Delta$=31 and 18 meV; $\Delta_p$= 22 and
15 meV. Left inset: fits of SIN tunneling spectra measured on
underdoped Bi-2212 \cite{27} by using three-gap model, with
$\Delta_{p}$=44, 28 and 20 meV and the set of gap values
$\Delta$=38, 26 and 17 meV for 46.4 K, $\Delta$=37, 25 and 16 meV
for 63.3 K and $\Delta$=36, 24 and 15 meV for 76 K. Right inset: fit
of SIN tunneling spectrum measured on slightly underdoped Bi-2212 at
50 K \cite{28} by using three gap model, with $\Delta$=36, 24 and 15
meV; $\Delta_p$= 73, 56 and 39 meV.}
\end{figure}
\begin{figure}
\includegraphics[width=0.4\textwidth]{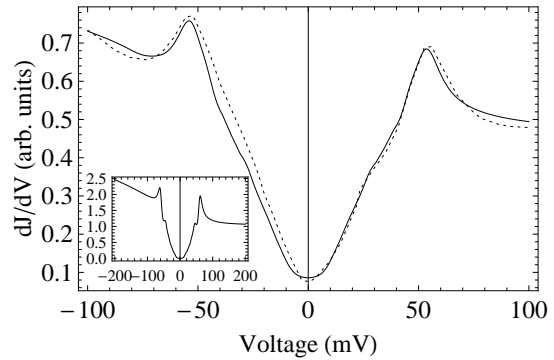}
\caption{\label{fig4:epsart}\footnotesize Comparison of the
tunneling conductance data on inhomogeneous Bi-2212 (dashed line)
\cite{23} with the tunneling conductance calculated at 30 K  (main
panel) using the multi-gap model ($\Delta$=52, 45, 36, 27, 19 and 12
meV; $\Delta_p$= 78, 65, 47, 34, 23 and 15 meV). Inset represents
the conductance curve with pronounced shoulder-like feature,
calculated at 30 K using the multi-gap model ($\Delta$=59, 43, 34,
25 and 16 meV; $\Delta_p$= 64, 52, 44, 28 and 18 meV).}
\end{figure}
We obtained the best fits to the experimental spectra by taking only
two or three terms in Eqs.(\ref{Eq.7}) and (\ref{Eq.8}). In this
way, we succeeded in fitting almost all of experimental conductance
curves by taking different gap values. The V-shaped subgap feature,
the asymmetric peaks and the dip-hump features, their temperature
dependences observed in tunneling spectra of underdoped Bi-2212
(left inset in Fig.3), slightly underdoped Bi-2212 (right inset in
Fig.3) and overdoped Bi-2212 (main panel in Fig.3) are well
reproduced. Moreover, the multi-gap model reproduces other tunneling
spectra (with varying local gap value, ranging from 20 to 70 meV)
and shoulders inside the conductance peaks observed in Bi-2212
\cite{23,24}. In particular, this model reproduces rather well one
of the experimental spectra of inhomogeneous Bi-2212 \cite{23} by
taking six terms in Eqs.(\ref{Eq.7}) and (\ref{Eq.8}), as shown in
Fig.4 (main panel). Further, the conductance curve calculated using
the multi-gap model (inset in Fig.4) is similar to that in Fig.1{\it
c} of Ref. \cite{24} measured on inhomogeneous Bi-2212.

We now discuss the relation between the BCS tunneling gap and the SC
order parameter. The unusually large reduced-gap values
$2\Delta/k_BT_c\simeq7-22$ observed in Bi-2212 \cite{15} compared to
the BCS value 3.52 give evidence that the BCS gap determined by
tunneling and ARPES measurements does not close at $T_c$ and it is
not related to the SC order parameter. While the peak suppression on
the negative bias side near $T_c$ observed in Bi-2212 is due to a
spectral superposition of the tunneling conductances associated with
the BCS DOS and quasi-free state DOS (originating from the polaron
dissociation). The persistence of the conductance peak on the
positive bias side well above $T_c$ is evidence for the opening of a
non-SC BCS gap at $T^*$ (for which the ratio $2\Delta/k_BT^*$
remains constant and close to the value 3.52 \cite{1,2,3}). The
pre-formed Cooper pairs condense into a SF Bose-liquid state at
$T_c$ (at which the SC order parameter appears) and the BCS pairing
gap persists as the non-SC gap both below $T_c$ and above $T_c$
\cite{1,2}. The optical measurements (including tunneling
spectroscopy and ARPES) are mainly sensitive to the excitation gaps
at $\varepsilon_F$, but such experimental probes compared with the
thermodynamic methods \cite{42,43} and the methods of critical
magnetic field \cite{44} and current \cite{45} measurements are
insensitive to the identification of the SC order parameter as the
SF condensation energy or as the energy needed for destruction of a
SF Bose-condensate (see also Refs. \cite{1,2,46}).

{\it Conclusion.}-- We have proposed a model describing the
distinctive mechanisms of quasiparticle tunneling across the SIN
junction at negative and positive biases. The model incorporating
effects of the BCS DOS and quasi-free state DOS (appearing at the
polaron dissociation) at negative bias, and the gap inhomogeneity
(i.e., multi-gap effects) reproduces the nearly U- and V-shaped and
shoulder-like subgap features, peak-dip-hump structure and asymmetry
of the conductance peaks and their evolution with temperature and
doping as seen in tunneling spectra of Bi-2212. In this model, many
unusual features of the tunneling spectra observed in Bi-2212 on the
negative bias side arise from the spectral superposition of the
tunneling conductances associated with the BCS DOS, quasi-free state
DOS and multi-channel tunneling.

We thank E.M. Ibragimova, B.L. Oksengendler, P.J. Baimatov, B.Y.
Yavidov and B.V. Turimov for useful discussions. This work was
supported by the Foundation of Uzbek Academy of Sciences, Grant No.
FA-F2-F070+075.

\end{document}